\documentclass[twoside,reqno]{HERON}
\usepackage{epsfig,cite,colordvi}
\usepackage{graphicx}
\usepackage{url}
\usepackage{amssymb,amsmath,amscd,epsf}
\usepackage{times}
\usepackage{makeidx}
\begin{document}

\title{Effects of Short-Range Correlations on Quarkyonic Matter}

\runningheads{Effects of Nuclear Symmetry Energy on Neutron Dark Decay in Neutron Stars}{K. Folias, Ch.C. Moustakidis}

\begin{start}

\coauthor{K. Folias}{ }, \author{Ch.C. Moustakidis}{}

\index{Author, F.A.}
\index{Author Jr., S.B.}
\index{Author, C.}

\address{Department of Theoretical Physics, Aristotle University of Thessaloniki, 54124 Thessaloniki, Greece}{}


\begin{Abstract}
In this work, we focus on investigating the impact of the short-range correlations on the properties of quarkyonic matter. These correlations induce the scattering of nucleons to momenta exceeding the Fermi momentum, thereby altering the momentum-space occupation distribution that underlies the quarkyonic framework.   In particular,  we examine to what extent the emergence of a high-momentum tail, due to the short-range nucleon-nucleon correlations,  in the momentum distribution, for momenta exceeding the Fermi momentum, affects the fundamental properties of quarkyonic matter, such as the energy density, pressure, and speed of sound, among others. 
In fact, at high densities, the behavior of the momentum distribution has been established, both experimentally and theoretically, to follow a characteristic $\sim 1/k^4$
dependence at large momenta.
To the best of our knowledge, this effect on quarkyonic matter has not yet been systematically investigated. Moreover, this experimentally confirmed feature of the momentum distribution suggests that heavy-ion collisions may probe quarkyonic matter, underscoring the importance of incorporating short-range correlations into a complete quarkyonic model.

\end{Abstract}
\end{start}

\section{Introduction}
One of the longstanding open problems in modern Nuclear Physics and Astrophysics concerns the composition and properties of dense nuclear matter, both at zero and at finite temperature~\cite{Glendenning1996CompactSN,Shapiro:1983du,Haensel2007NeutronS1,schaffner-bielich_2020}. A detailed understanding of this issue is of fundamental importance for the study of phenomena observed in heavy-ion collision experiments performed in terrestrial laboratories, as well as for achieving a deeper insight into the internal structure and physical properties of neutron stars. The theoretical scenarios considered so far include the conventional description of baryonic matter composed of neutrons and protons. Extensions of this framework also allow for the presence of various hyperon species, as well as the possible condensation of pions and other mesonic degrees of freedom. Another widely discussed possibility is the appearance of deconfined quark matter in the inner core of neutron stars, arising through different proposed phase–transition mechanisms. The study of such hybrid stars, and more generally the attempt to reconcile the simultaneous presence of baryonic and quark degrees of freedom, has also led to the proposal of a hybrid phase known as quarkyonic matter, which has attracted considerable attention in recent years~\cite{Baym-2019}.
An interesting attempt in this direction is to consider a hybrid state of dense nuclear matter called quarkyonic matter~\cite{PhysRevC.110.025201,PhysRevC.107.065201,Cao:2022inx,Kovensky:2020xif,Koch_2023,Jeong:2019lhv,PhysRevLett.122.122701,McLerran:2020rnw,article,Fukushima_2016,PhysRevC.108.045202,PhysRevLett.132.112701,McLerran-2024,Bluhm-2024,Lysenko-2025,Gao-2025a,Gao-2025b,Pang-2024,Yamamoto-2023,Ciao-2020,Sen-2021a,Ivanytskyi-2025,Zhao-2020,Margueron-2021,Kojo-2021,Folias-2025,Folias-2026,Gartlein-2025,Sen-2021,Xia-2023,Bluhm-2025}. In this consideration, quarkyonic matter both quarks and nucleons appear as quasiparticles. According to the analysis of Refs.~\cite{PhysRevLett.122.122701,McLerran:2020rnw,article}, the basic assumption of quarkyonic matter  is that at large Fermi energy, the degrees of freedom inside the Fermi
sea may be treated as quarks, and confining forces remain
important only near the Fermi surface where nucleons
emerge through correlations between quarks. In quarkyonic matter, confinement at the Fermi surface
produces triplets with spin $1/2$ that we identify with
baryons that  occupy  a momentum shell of width $\Delta \simeq \Lambda_{QCD}$ (where $\Lambda_{QCD}$ represents  the confinement
scale)~ \cite{PhysRevLett.122.122701,Fukushima_2016,PhysRevC.108.045202,PhysRevLett.132.112701,McLerran-2024}. The width of the momentum shell $\Delta$ depends on the baryon density. 
Short-range correlations (SRCs) generate a high-momentum tail in the nucleon momentum distribution (MD), leading to a non-negligible fraction of nucleons occupying states above the Fermi momentum~\cite{Antonov-1988,Flynn-1984,Stringari-1990,Migdal-1957,Galitski-58a,Galitski-58b,Czyz-1961,Belyakov-1961,Sartor-1981,Krotscheck-2020}. This feature, characterized by an approximately universal $\sim 1/k^4$ behavior, depends on the density and isospin composition of nuclear matter and can significantly influence on the  equation of state nuclear matter and neutron star matter (see the recent review~\cite{Bao-2026} and also the relevant Refs.~\cite{Lourenco-2026,Wang-2024,Souza-2020,Gautam-2025,Liu-2024}). In the present work, we incorporate the effects of SRCs into the quarkyonic matter framework by allowing a fraction of neutrons to occupy momentum states above the Fermi momentum. To describe the modified neutron MD, we adopt the phenomenological approach developed in Refs. \cite{Cai-2015,Cai-2016,Gautam-2025,Lu-2021,Lu-2022}, which provides a systematic prescription for introducing the high-momentum tail. This extension enables a consistent investigation of the role of SRCs in the description of quarkyonic matter and their implications for its thermodynamic properties. Our approach emphasizes the importance of accounting for high-momentum nucleon states within the quarkyonic framework, providing a more comprehensive description of dense nuclear matter and a basis for assessing the quantitative impact of SRCs.

The paper is organized as follows. Section~\ref{Themodel} present the theoretical framework. Section~\ref{Results} presents and discusses the results, while Section~\ref{Concl} summarizes the main conclusions.

\section{The theoretical model}
\label{Themodel}
\subsection{Momentum distribution of neutron matter}

We consider the MD  of neutrons $f(k,k_{F_n})\equiv n(k)$, modified by the effects of SRCs, to be described by the following ansatz~\cite{Cai-2015,Cai-2016,Lu-2021,Lu-2022}
\begin{eqnarray}
\hspace{-0.4cm}
n(k)&=&\left\{
		\begin{array}{ll}
			D_n+b_nI(|{\bf k}|/k_{F_n}), \  0 \leq {\bf k}\leq k_{F_n} & \\
			\\
			C_n\left(k_{F_n}/{\bf |k|} \right)^4, \  k_{F_n} \leq |{\bf k}| \leq \lambda_n k_{F_n} & \
            \\
			\\
			0, \   k \geq \lambda_n k_{F_n} & \
		\end{array}
	\right.
	\label{Bao-1}
\end{eqnarray}
where the Fermi momentum is given by  $k_{F_n}=(3\pi^2 n_n)^{1/3}$.
In Eq.~(\ref{Bao-1}), the  factor $D_n$ measures the depletion of the Fermi sphere
at zero momentum with respect to the Free Fermi gas model  model while $b_n$
is the strength of the momentum dependence $I(|{\bf k}|/k_{F_n})$  of the
depletion near the Fermi surface.
For the function $I(|{\bf k}|/k_{F_n})$ we consider the quadratic suggestion $I(|{\bf k}|/k_{F_n})=(|{\bf k}|/k_{F_n})^2$~\cite{Cai-2015,Cai-2016}. The normalization condition of the MD is 
\begin{equation}
2/(2\pi)^3\int n(k)d^3k=n_n
\label{norm-nn-1}
\end{equation}
The above normalization condition allows us to determine the parameter $D_n$ through the expression
\begin{equation}
D_n=1-\frac{3\beta_n}{k_{F_n}^3}  \int_0^{k_F^n} I(|{\bf k}|/k_{F_n})k^2dk-3C_n\left(1-\frac{1}{\lambda_n}  \right)
\label{Deltan-1}
\end{equation}
Moreover, the jump $Z_F^n$ of the MD at $k_{F_n}$ is defined as  $Z_F^n=n(k_{F_n-0})-n(k_{F_n+0})$ is associated with the strength of the short-range interactions and, in this sense, also serves as a quantitative measure thereof. In the present model is given by $Z_F^n=D_n+b_n-C_n$.


\subsection{Nuclear model for pure neutron matter}

We use the neutron Fermi momentum, $k_{F_n}$, as the key variable in our calculations rather than the neutron number density, $n_n$.
As a first approximation, assuming that nuclear matter consists entirely of neutrons, the energy density is given by
\begin{equation}{\cal E}_{n}(k_{F_n}) =E_{\rm kin}(k_{F_n})+mc^2 n_n+V_{\rm int}(k_{F_n})
\label{eq2}
\end{equation}
where the kinetic energy density is given by 
\begin{equation}
E_{\rm kin}(k_{F_n})=\frac{g_s}{(2\pi)^3}\int \frac{\hbar^2k^2}{2m}f_n(n_n,k) d^3k   
\end{equation}
The potential term $V_{\rm int}(k_{F_n})$, in the present study,  depends not only on the density of neutrons but also on their momentum. In particular, we use a  model  which has been extensively applied in the literature and has the form (see Refs.~\cite{Prakash-1997,Bertsch-1988,Gale-1987,Li-2008,Moustakidis-2007,Moustakidis-2008,Moustakidis-2009a,Moustakidis-2009b} and references therein)
 \begin{eqnarray}
V_{\rm int}(k_{F_n})&=&\frac{1}{3}An_0\left(1-x_0   \right)u^2 +\frac{\frac{2}{3}Bn_0\left(1-x_3\right)u^{\sigma+1} }{1+\frac{2}{3}B'\left(1-x_3   \right)u^{\sigma-1}}\nonumber\\
&+& u\sum_{i=1,2}\frac{1}{5}\left[\frac{}{}6C_i-8Z_i \right]{\cal J}_n^i(k_{F_n})
\label{eq5}
\end{eqnarray}
where 
\begin{eqnarray}
{\cal J}_{n}^i(k_{F_n})&=&\frac{2}{(2\pi)^3}\int{\rm g}(k,\Lambda_i) f_n(n_n,k) d^3k
\label{Jn-1}
\end{eqnarray}
We use the notations $u =n_n/n_0$ (where $n_0$ is  the saturation density $n_0 = 0.16$ fm$^{-3} $),  $\Lambda_1 = 1.5k_{F{n_0}}$, $\Lambda_2 = 3k_{F{n_0}}$ (where  $k_{F{n_0}}$ is the neutron Fermi momentum at the saturation density $n_0$). 
The parameterization of  Eq.~(\ref{eq5}) is given in Ref.~\cite{Prakash-1997}.
The parameters $A$, $B$, $B'$ $\sigma$, $C_1$ and $C_2$, which
appear in the description of symmetric nuclear matter and the
additional parameters $x_0$, $x_3$, $Z_1$, and $Z_2$ used to determine
the properties of asymmetric nuclear matter, are treated as
parameters constrained by empirical knowledge.
The first two terms of the right-hand side of Eq.~(\ref{eq5}) arise
from local contact nuclear interactions that led to power density
contributions such as in the standard Skyrme equation of state.
Moreover,  the first term  represents an attractive interaction, while the second one  corresponds to a repulsive interaction that becomes dominant at high densities (for $n > 0.6$ fm$^{-3}$). 
The third term describes the
effects of finite-range interactions according to the chosen
function ${\rm g}(k,\Lambda_i)$ which, in the present work has the form~\cite{Prakash-1997,Bertsch-1988,Gale-1987}
\begin{equation}
{\rm g}(k,\Lambda_i)=\left[1+\left(\frac{k}{\Lambda_i}  \right)^2 \right]^{-1}  
\label{g-1}
\end{equation}
The  fraction $\chi_n^{HMT}$ of the neutrons in the high momentum  tail  is given by   
\begin{equation}
\chi_n^{HMT}=3C_n\left(1-\frac{1}{\lambda}\right)
\label{fraction-hmt}
\end{equation}
By appropriately adjusting the parameters $\lambda$, $C_n$, and $b_n$, we can obtain the desired value of the high-momentum fraction $\chi_n^{\mathrm{HMT}}$ and, consequently, the corresponding value of the momentum jump $Z_F^n$. In the present study, we set $\chi_n^{\mathrm{HMT}}=0.1$. However, the analysis can readily be extended to a broader range of values consistent with the corresponding experimental estimates.

\subsection{The quarkyonic model} 
The chargeless two-flavor quarkyonic model assumes that strongly interacting quarks near the Fermi surface form interacting neutrons, while the remaining up (u) and down (d) quarks are noninteracting and occupy the lowest-momentum states up to their respective Fermi momenta, $k_{_Fu}$ and $k_{F_d}$~\cite{PhysRevLett.122.122701,Zhao-2020}.
The total baryon density is given by
\begin{equation} 
n_B = n_n + \frac{(n_u + n_d)}{3} \label{eq21}
\end{equation}
with due to chemical equilibrium 
\begin{equation}
\mu_n=\mu_u+2\mu_d
\label{chemi-eq-1}
\end{equation}
The quark chemical potentials will be given by
\begin{equation}\mu_i=\sqrt{(\hbar c k_{F_i})^2+ m_{Q}^2c^4}, \qquad i=u,d 
\label{eq13}
\end{equation}
In this case the lower value of the momentum for the neutrons is defined as $k_{0n}$
\begin{equation}
 k_{0n}=(k_{F_n}-k_{tn})\left[1+\frac{\Lambda^2}{(\hbar c)^2 k_{F_n}k_{tn}}  \right] 
 \label{kon-1}
\end{equation}
where the Fermi momenta transition $k_{tn}$ is obtained from beta equilibrium of the uniform 
n,u,d system at the transition density $n_{tn}$.
Moreover, the width of the momentum shell, occupied by neutrons $\Delta=(k_{F_n}-k_{0n})$,  is given by
\begin{equation}
 \Delta=k_{tn}-\frac{\Lambda^2(k_{F_n}-k_{tn})}{(\hbar c)^2 k_{F_n}k_{tn}}  
 \label{With-1}
\end{equation}
In the qyarkyonic model, conserning the neutron matter the corresponding quantities, given in Eqs.~(\ref{eq2})-(\ref{Jn-1}) are recalculated, with the limits of integration  $k_{0n} \leq {\bf k}\leq k_{F_n}$. The quark energy densities are given by the expressions 
for noninteracting fermions 
\begin{equation}{\cal E}_{Q}(n_i,T) = \frac{g_s N_c}{2\pi^2}\sum_{i=u,d} \int_0^{h_{i}}  \sqrt{(\hbar c k)^2+ m_{i}^2c^4} k^2 dk
\label{eq16}
\end{equation}
Quark masses are obtained by $m_Q = m_N/N_c$ where $m_N$ is the nucleon mass and $N_c$ is the number of colors. We set the number of colors and the degeneracy of the spin equal to $N_c = 3 $ and $g_s = 2$ respectively, and we introduce the quarkyonic matter at a baryon density around $n_B = 0.2 - 0.4$ fm$^{-3}$. 
The fraction of the neutron in the high momentum $\chi_n^{HMT}$ tail is given by
\begin{equation}
\chi_n^{HMT}=\frac{3C_n k_{F_n}^3}{k_{F_n}^3-k_{0n}^3}\left(1-\frac{1}{\lambda_n}\right) 
\label{fraction-hmt-2}
\end{equation}
\subsection{The neutron star matter}

The total energy density, of neutron star matter,  is given by
\begin{equation} {\cal E}_{\rm tot}= {\cal E}_n + {\cal E}_Q 
\label{eq22}
\end{equation}
while the chemical potential for each species of matter is defined as 
\begin{equation} \mu_i = \frac{\partial {\cal E}_{\rm tot}}{\partial n_i}, \qquad i=n,u,d 
\label{eq23}
\end{equation}
Finally, the total pressure and the speed of sound are  given respectively  by
\begin{equation} 
P_{\rm tot} = - {\cal E}_{\rm tot} +  \sum_{i = n,u,d}\mu_i n_i\hspace{3mm}, \label{eq24}
\end{equation}
\begin{equation}
\frac{c_s}{c}=\sqrt{\frac{\partial P_{\rm tot}}{\partial {\cal E}_{\rm tot}}}
\label{speed-2}  
\end{equation}
The quantities $n_B$, ${\cal E}_{\rm tot}$, $P_{\rm tot}$ are functions of the momentum $k_{F_n}$. In totally, from Eqs.~(\ref{eq22})-(\ref{speed-2}),  the equation of state and the speed of sound  for the quarkyonic matter can be readily determined. The final step involves solving the Tolman–Oppenheimer–Volkoff equations~\cite{Glendenning1996CompactSN,Shapiro:1983du,Haensel2007NeutronS1,schaffner-bielich_2020} and calculating the bulk properties of neutron stars, such as their mass and corresponding radius.

\section{Results and Discussion}
\label{Results}
In the present study, we parameterize the MD  so as to ensure that, in both pure neutron matter and quarkyonic matter, $10\%$ of the neutrons occupy states with momenta above the Fermi momentum. Our primary objective is to investigate the effects of SRCs on the EoS of both pure neutron matter and quarkyonic matter, the latter being considered as the constituent matter of a neutron star. Accordingly, we consider four distinct cases: pure neutron matter without and with SRCs, and quarkyonic matter without and with SRCs. This approach allows us to systematically investigate all four scenarios and assess the impact of short-range correlations on the equation of state of each type of matter.

\begin{figure*}[ht]
 \centering
\includegraphics[width=0.47\textwidth]{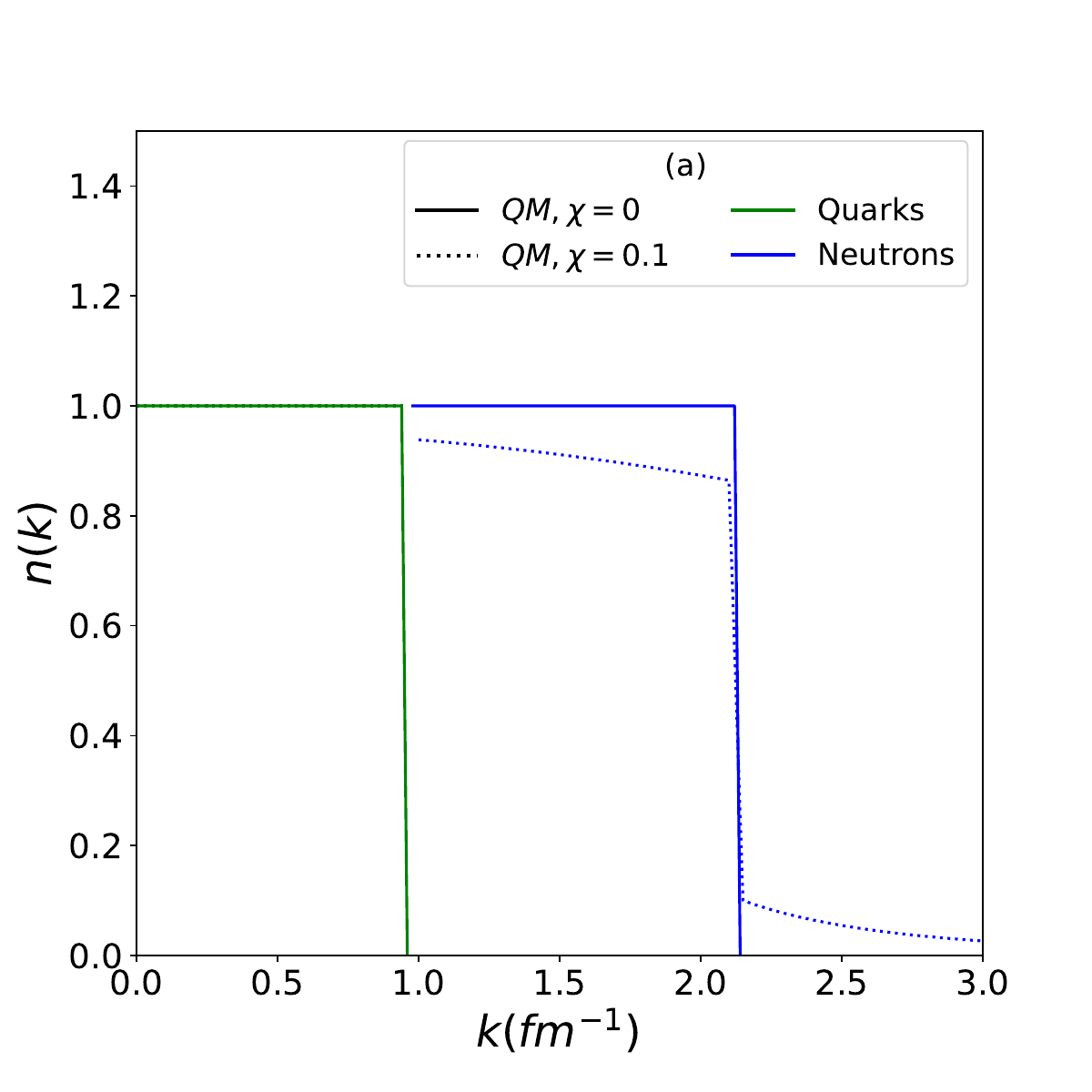}
\includegraphics[width=0.47\textwidth]{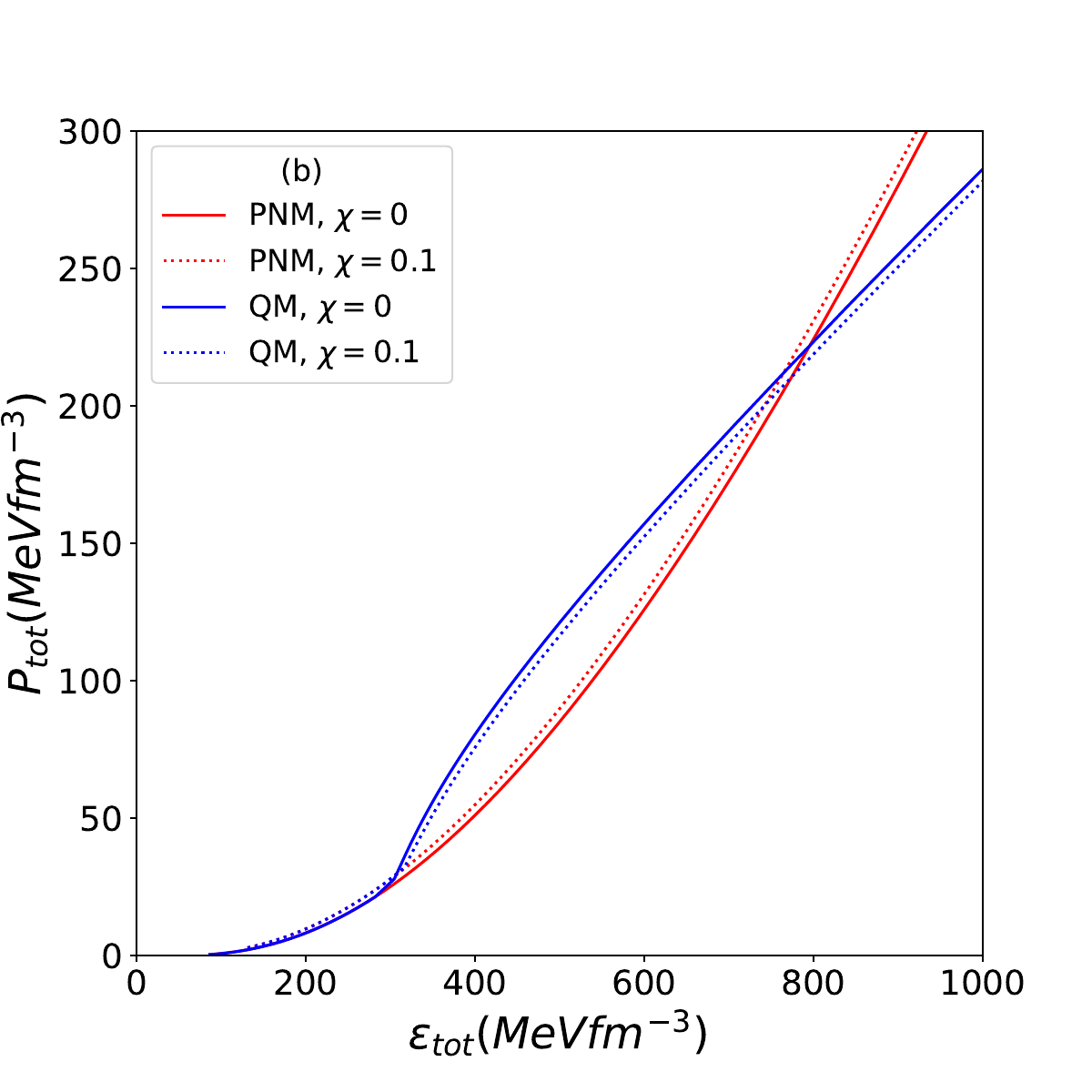}
\includegraphics[width=0.47\textwidth]{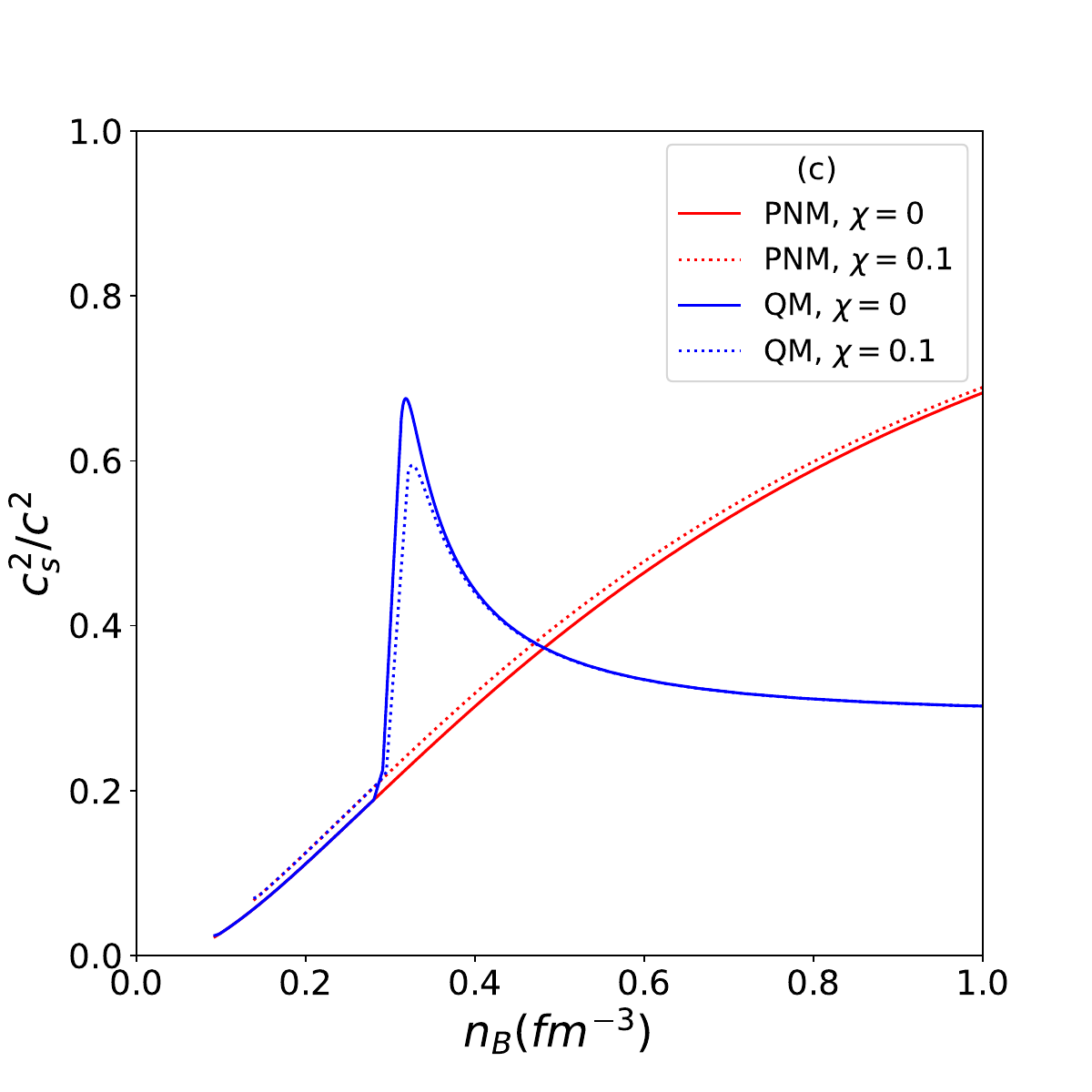} 
\includegraphics[width=0.47\textwidth]{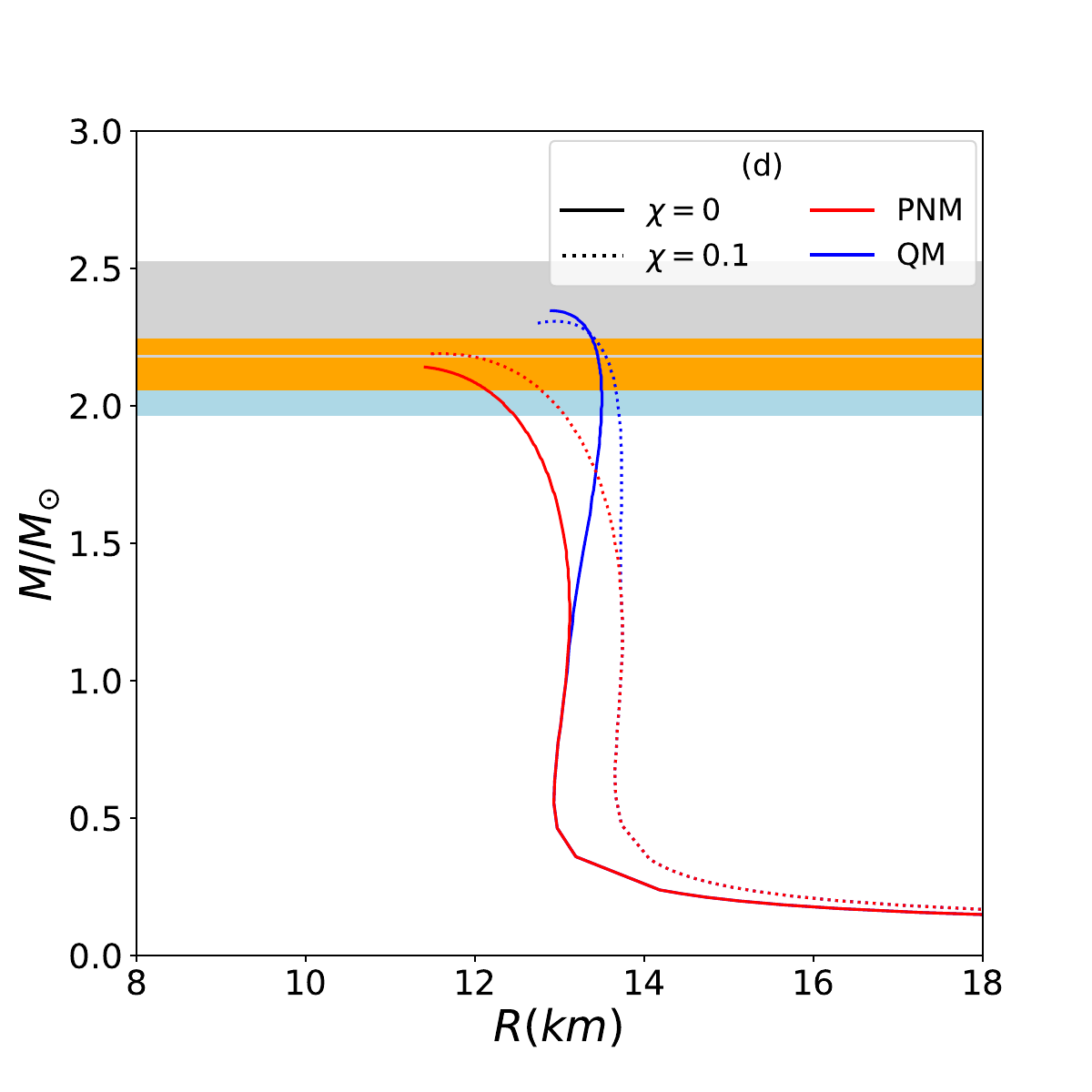}
\caption{(a) Momentum distributions of quarks and neutrons, assuming that $10\%$  ($\chi=0.1$) of the neutrons occupy momentum states above the Fermi momentum compared with the corresponding case in which the occupation is restricted to states within the Fermi sea ($\chi=0$). (b) Equations of state for the four cases considered. (c) Square of the speed of sound, expressed in units of the speed of light squared, as a function of baryon number density for the four cases considered. (d) Mass–radius relations for neutron stars corresponding to the four cases considered. }
  \label{M-R-L-z50}
\end{figure*}

Figure 1(a) illustrates the neutron MD with and without the inclusion of SRCs. The redistribution of neutrons toward momentum states above the Fermi momentum is clearly evident. It is worth emphasizing that this behavior occurs exclusively for neutrons, whereas, in quarkyonic matter, the quarks occupy momentum states up to their respective Fermi momenta, without any redistribution into states with higher momenta.  Figure 1(b) shows the four corresponding equations of state. It is evident that the presence of quarkyonic matter leads to a stiffer EoS, whereas the effect of SRCs is comparatively smaller, yet still non-negligible. An interesting finding is that SRCs have opposite effects on the stiffness of the equation of state  in the two cases considered. While SRCs lead to a stiffer EoS in pure neutron matter, they result in a slightly softer EoS in quarkyonic matter.

Figure 1(c) highlights the characteristic behavior of the speed of sound in quarkyonic matter, which exhibits a pronounced peak at relatively low densities before approaching, at higher densities, the expected asymptotic value characteristic of deconfined quark matter. In this case, as well, the effects of SRCs are relatively small but not negligible.  Figure 1(d) shows the stellar mass as a function of radius, reflecting the behavior of the underlying EoS. More specifically, at relatively low densities, the EoSs incorporating SRCs are stiffer than those without SRCs. At higher densities, however, the quarkyonic matter EoS exhibits a softening when SRCs are included. Consequently, at lower stellar masses, quarkyonic stars with SRCs have comparatively larger radii than their counterparts without SRCs, whereas the corresponding radii tend to converge as the stellar mass increases.

\section{Conclusions}
\label{Concl}
To the best of our knowledge, the effects of SRCs on quarkyonic matter are investigated for the first time in the present work. Assuming that $10\%$ of the neutrons occupy momentum states above the Fermi momentum, we examine the extent to which this redistribution affects both the corresponding EoS and the fundamental properties of neutron stars, particularly their masses and radii. Our results demonstrate that SRCs modify the behavior of the EoS, leading to larger stellar radii while leaving the maximum mass nearly unchanged. These findings indicate that SRCs should be taken into account in models of quarkyonic matter in neutron stars, as their effects are physically significant and cannot be neglected. A natural next step is to investigate their impact on neutron-star matter in beta equilibrium, where an even greater depletion of the Fermi sea is expected.

\end{document}